\documentstyle[multicol, aps, epsf]{revtex}

\begin{document}
\twocolumn[\hsize\textwidth\columnwidth\hsize\csname@twocolumnfalse%
\endcsname
\draft

\title
{\large \bf Comment on ``Relating atomic-scale electronic
phenomena to wave-like quasiparticle states in superconducting
Bi$_2$Sr$_2$CaCu$_2$O$_{8+\delta}$''}
\author{Khee-Kyun Voo$^1$, Hong-Yi Chen$^2$, and Wen-Chin Wu$^3$}
\address{$^1$ Department of Physics, National Tsing-Hua University,
Hsinchu 30043, Taiwan}
\address{$^2$ Texas Center for Superconductivity and Department of
Physics, University of Houston, Houston, TX 77204, USA}
\address{$^3$ Department of Physics, National Taiwan Normal University,
Taipei 11650, Taiwan}

\maketitle

\begin{abstract}

\end{abstract}

]

In a recent article \cite{M02}, McElroy $et~al.$ argued that the
peaks observed by scanning tunneling measurements in the
reciprocal space of the spatial modulation of local
density-of-states (LDOS) in superconducting
Bi$_2$Sr$_2$CaCu$_2$O$_8$ can be understood as due to
quasiparticle interference between regions of high DOS on the
Fermi surface. They proposed an ``octet'' model and have managed
to describe many essential features of the peaks. In their paper,
they pointed out that the location of the peaks possess mirror
symmetry about the lattice axes and diagonals. In this Comment, we
point out that there should exist a further symmetry, the
$umklapp$ symmetry in a periodic lattice. As a consequence, some
extra peaks which were not mentioned by the authors, should exist
within their model and may have escaped from being detected.
Moreover, these peaks can also disrupt the distinguishability of
individual peaks.

In Fig.~1, we have illustratively shown some direct and $umklapp$
scattering wavevectors in the octet model at a bias voltage. The
location of the wavevectors has now additional reflection
symmetries about $q_x, q_y = \pm \pi, \pm 3\pi, \pm 5\pi, etc$.
There are two kinds of nonequivalent $umklapp$ generated peaks.
Focusing on the quadrant $0 < q_x
< \pi$ and $0 < q_y < \pi$ (peaks in other quadrants are obtained
through symmetry operation on this), an extra peak denoted by
$q_5'$ is seen on the lattice axis when the bias voltage exceeds a
threshold. Peak $q_5$ was reported in Ref. 1 but $q_5'$ was not.
Since $umklapp$ symmetry is dictated by the lattice and estabished
by say, angle-resolved photoemission spectroscopy, $q_5'$ should
exist correspondingly. More disturbingly, there also exists an
$umklapp$ partner of $q_4$ (denoted by $q_4'$) which may touch
$q_{2,6}$, or vice versa speaking. We draw the attention of the
readers to the fact that it was mentioned in Ref.~\cite{M02} that
$q_4$ had been hard to detect, and we propose that this may be due
to the close neighboring of two peaks.

In conclusion, there should exist $umklapp$ peaks to complete the
quasiparticle interference picture, and those peaks may already
have disrupted current data analysis. The crowding in of the
$umklapp$ peaks into the quadrant can make the peaks hard to
discern above some bias voltage. Raising the temperature at a
fixed bias voltage will also be detrimental to the peaks, since
the superconducting gap is closed and as a result the umklapp
peaks crowd in. This could be part of the reason of the observed
degradation of those peaks at higher biases mentioned in
Ref.~\cite{M02}.

%
%
%
%
%
%
%


\begin{figure}[t]
\mbox{ \epsfxsize=0.95\hsize{\epsfbox{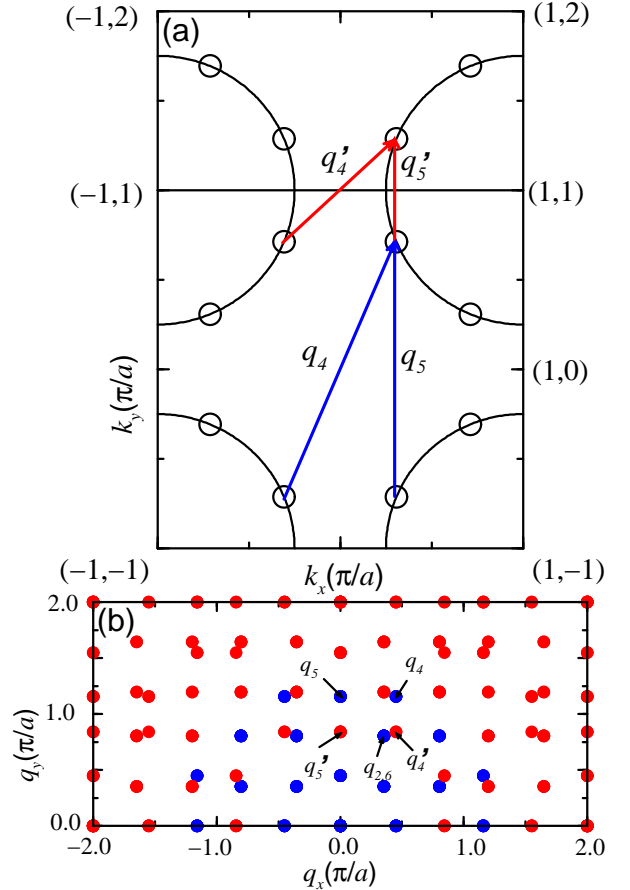}}} \vspace{0.5cm}
\caption{(a) Some scattering wavevectors due to the octet model.
Vectors discussed by McElroy $et$ $al.$ [1] are shown in blue, and
the $umklapp$ vectors are in red. (b) Location of all the
wavevectors. The $umklapp$ vectors are in red.} \label{fig}
\end{figure}


\end{document}